# Fully strained epitaxial Ti$_{1-x}$Mg$_x$N(001) layers

Baiwei Wang and Daniel Gall

*Department of Materials Science and Engineering, Rensselaer Polytechnic Institute, Troy, NY 12180, USA*

Ti$_{1-x}$Mg$_x$N(001) layers with $0.00 \leq x \leq 0.49$ are deposited on MgO(001) by reactive magnetron co-sputtering from titanium and magnesium targets in 5 mTorr pure N$_2$ at 600 °C. X-ray diffraction $\omega$-$2\theta$ scans, $\omega$-rocking curves, $\varphi$-scans, and high resolution reciprocal space maps show that the Ti$_{1-x}$Mg$_x$N layers are rock-salt structure single crystals with a cube-on-cube epitaxial relationship with the substrates: (001)$_{TiMgN}$ ‖ (001)$_{MgO}$ and [100]$_{TiMgN}$ ‖ [100]$_{MgO}$. Layers with thickness $d$ = 35-58 nm are fully strained, with an in-plane lattice parameter $a_\parallel$ = 4.212 ± 0.001 Å matching that of the MgO substrate, while the out-of-plane lattice parameter $a_\perp$ increases with $x$ from 4.251 Å for TiN(001) to 4.289 Å for Ti$_{0.51}$Mg$_{0.49}$N(001). This yields a relaxed lattice parameter for Ti$_{1-x}$Mg$_x$N(001) of $a_o$ = (1-$x$)$a_{TiN}$ + $xa_{MgN}$ − $bx$(1-$x$), where $a_{TiN}$ = 4.239 Å, $a_{MgN}$ = 4.345 Å, and the bowing parameter $b$ = 0.113 Å. In contrast, thicker Ti$_{1-x}$Mg$_x$N(001) layers with $d$ = 110-275 nm are partially (pure TiN) or fully ($x$ = 0.37 and 0.39) relaxed, indicating a critical thickness for relaxation of 50-100 nm. The in-plane x-ray coherence length is large (100-400 nm) for fully strained layers with $0.00 \leq x \leq 0.45$ but drops by an order of magnitude for $x$ = 0.49 as the composition approaches the phase stability limit. It is also an order of magnitude smaller for thicker ($d \geq$ 110 nm) layers, which is attributed to strain relaxation through the nucleation and growth of misfit dislocations facilitated by glide of threading dislocations.

## I. Introduction

Titanium magnesium nitride Ti$_{1-x}$Mg$_x$N has gained recent interest because of two reasons: Firstly, it is a promising material for hard wear-protective [1, 2], optical decorative [3, 4], and bioimplant [5] coatings, as it shares the prominent physical properties of traditional transition metal nitrides including high hardness, wear and corrosion resistance, and high temperature stability [6-11]. Secondly, the stoichiometric Ti$_{0.5}$Mg$_{0.5}$N solid solution is a semiconductor with a predicted 1.3 eV band gap [12], a measured negative temperature coefficient of resistivity [13] and a negligible density of states at the Fermi level [14]. Therefore, Ti$_{1-x}$Mg$_x$N layers with $x \leq$ 0.5 have a tunable infrared plasmonic activity [15, 16] and are promising as, for example, plasmonic solar heat transducers [17]. We envision that all these applications will benefit from a better understanding of the Ti$_{1-x}$Mg$_x$N growth process, including the interplay between composition, strain, and stress relaxation. Studying the epitaxial growth process has the potential to provide such understanding as it is a model system that effectively eliminates the confounding effects from crystal orientation variations and grain boundaries that are present in polycrystalline samples [18]. In addition, epitaxial Ti$_{1-x}$Mg$_x$N(001) layers are also promising for plasmonic and semiconductor applications as they minimize electronic losses due to crystalline defects that deteriorate the plasmonic field enhancement [15] and cause carrier scattering and mid-gap traps for the predicted Ti$_{0.5}$Mg$_{0.5}$N semiconductor [12]. Reactive magnetron sputtering in a N$_2$ atmosphere is the most common approach for the deposition of epitaxial transition metal nitride layers, and has been successfully employed to grow TiN(001) [8, 19-22], ScN(001) [21, 23-25], CrN(001) [26-28], NbN(001) [29], MoN(001) [30], HfN(001) [31, 32], TaN(001) [33-36], WN(001) [37, 38], and CeN(001) [39], as well as ternaries like $Ti_{1-x}W_xN$(001) [40], $Ti_{1-x}Sc_xN$(001) [41], $Sc_{1-x}Al_xN$(001) [42-44], and $NbC_{1-x}N_x$(001) [45, 46]. These layers often exhibit residual stress, which is due to multiple contributing effects including lattice misfit with the substrate [19, 30, 38], energetic incident particles which may cause interstitial point defects and/or vacancies [47-49], differential layer/substrate thermal contraction after deposition in combination with a kinetic barrier for dislocation formation [45, 50], and atomic shadowing leading to voids [51, 52]. Stress/strain relaxation in thin films is a fundamental process that affects structural and morphological stabilities as well as electronic properties. It is therefore of significant interest for device applications as well as for protective coatings since the adhesive strength, hardness, fracture toughness, tribological properties, and resulting mechanical wear and failures are strongly influenced by strain [53, 54], while lattice strain in semiconductor and optoelectronic devices also provides opportunities to tune band gaps/alignments/degeneracy, band curvatures and hence carrier effective masses [55], but may also lead to dislocations and associated defect states that trap conduction electrons. Strain relaxation in epitaxial $Ti_{1-x}Mg_xN$(001) as a function of composition, layer thickness, and processing conditions is completely unexplored, which motivates the present study.

In this paper, we report on the deposition of single crystal $Ti_{1-x}Mg_xN$(001) layers ($0.00 \leq x \leq 0.49$) on MgO(001) by reactive magnetron co-sputtering from titanium and magnesium targets in 5 mTorr pure $N_2$ at 600 °C. X-ray diffraction $\omega$-$2\theta$ scans, $\omega$-rocking curves, $\varphi$-scans, and high resolution reciprocal space maps show that rock-salt $Ti_{1-x}Mg_xN$ layers are epitaxial single crystals. Layers with thickness 35 nm $\leq d \leq$ 58 nm are fully strained, while thicker layers with $d = 110 - 275$ nm are nearly ($x = 0$) or completely ($x \geq 0.37$) relaxed, suggesting that relaxation occurs above a critical thickness for the formation of misfit dislocations. The in-plane x-ray coherence length for fully strained layers with $0.00 \leq x \leq 0.45$ is 100-400 nm, but drops to 20 - 44 nm for relaxed layers. Similarly, the out-of-plane x-ray coherence length matches the layer thickness for strained layers but is reduced to 35 - 47 nm for relaxed layers, which is attributed to strain fields associated with misfit and threading dislocations. Increasing $x$ to 0.49 causes a decrease in both correlation lengths, which is attributed to local strain variations as the alloy approaches a composition for phase separation.

## II. Experimental Procedure

$Ti_{1-x}Mg_xN$ layers were deposited by reactive magnetron co-sputtering in a load-locked ultra-high vacuum deposition system with a base pressure of $10^{-9}$ Torr [42]. Double-side polished $10 \times 10 \times 0.5$ mm$^3$ single-crystal magnesium oxide MgO(001) substrates were cleaned in sequential ultrasonic baths of tri-chloroethylene, acetone, and isopropyl alcohol, for 20 min each, rinsed in de-ionized water, blown dry with nitrogen, mounted onto a Mo substrate holder using silver paint, and inserted into the deposition system. Prior to deposition, substrates were degassed for 1 hour at 1000 °C using a radiative pyrolytic graphite heater. Subsequently, the heater current was adjusted to reach the desired substrate temperature of 600 °C, as measured by a thermocouple underneath the substrate holder that was cross-calibrated with a pyrometer focused on the substrate surface. 99.999% pure $N_2$, which was further purified with a MicroTorr purifier, was introduced into the chamber with a needle valve to reach a constant pressure of 5 mTorr, as measured with a capacitance manometer. 5-cm-diamter nominally 99.99% pure Ti and Mg targets were positioned 9 and 23 cm from the substrate surface with 45° and 0° tilt angles, respectively. Before deposition, the Ti and Mg targets were sputter etched for 5 min using 100 W on each magnetron with a shutter shielding the substrate. The two magnetrons were simultaneously operated with separate DC power supplies, keeping the power to the Ti target constant at 100 W for all layers while varying the power to the Mg target from 0 to 350 W to achieve samples with different Ti-to-Mg compositions. The substrate was continuously rotated at 60 rpm to ensure composition and thickness uniformity. The potential to the Ti target (-455±4 V) was unaffected by the power to the Mg target, suggesting that the plasmas from the two deposition sources have a negligible interaction such that the Ti flux to the substrate is expected to be the same for all layers, corresponding to a constant TiN deposition rate of 1.4 nm/min. The deposition time for each sample was adjusted according to the Mg power, in order to obtain comparable film thicknesses of approximately 50 nm, as determined by X-ray reflectivity (XRR). A second set of $Ti_{1-x}Mg_xN$ layers with larger thicknesses of 110-275 nm were deposited in 15 mTorr $N_2$ for 1 hour, with Ti and Mg targets facing the substrate at the same distance of 9 cm and ±45° tilts as described in Ref. [13], keeping the power to the Ti target constant at 100 W to achieve a TiN deposition rate of 1.8 nm/min and adjusting the power to the Mg target from 0 to 200 W, resulting a set of samples with various compositions.

X-ray photoelectron spectroscopy (XPS) spectra were acquired using Al $K_\alpha$ radiation (1486.7 eV) in a PHI 5000 Versaprobe™ with a hemispherical analyzer and an 8-channel detector. The samples were analyzed after exposure to air but without any sputter cleaning of their surfaces, such that the XPS results are not affected by preferential sputtering [56, 57]. Rutherford Backscattering Spectroscopy (RBS) was done using 2 MeV $^4He^+$ ions incident at an angle of 6° relative to the sample surface normal, and a total scattering angle of 166.2° between incident beam and detector. RBS spectra were analyzed using the XRUMP simulation program.

X-ray diffraction was done with a Panalytical X'Pert PRO MPD system with a Cu $K\alpha$ source and a hybrid mirror with a two-bounce two-crystal Ge(220) monochromator, yielding a parallel incident beam with a wavelength $\lambda_{K\alpha 1}$ = 1.5406 Å, a divergence of 0.0068°, and a width of 0.3 mm. Sample alignment included height adjustment as well as correction of the $\omega$ and $\chi$ tilt angles by maximizing the substrate peak intensity. Symmetric $\omega$-$2\theta$ scans were obtained using a 0.04 radian Soller slit in front of a PIXcel solid-state line detector operated in receiving mode with a 0.165 mm active length, corresponding to a $2\theta$ opening of less than 0.04°. $\omega$-rocking curves were obtained using constant $2\theta$ angles corresponding to $Ti_{1-x}Mg_xN$ 002 reflections and using the same parallel beam geometry as used for $\omega$-$2\theta$ scans. Asymmetric high-resolution reciprocal space maps (HR RSM) around 113 reflections were obtained using a small angle (10-14°) between the sample surface and the diffracted beam to cause beam narrowing which increases the $2\theta$ resolution and therefore facilitates fast high-resolution reciprocal space mapping by taking advantage of parallel acquisition with all 255 channels of a line detector operated in scanning mode. XRD $\varphi$ scans were obtained using a point focus optics with a poly-capillary x-ray lens that provides a quasi-parallel Cu $K_\alpha$ beam with a divergence of less than 0.3° to minimize defocusing effects associated with the non-uniform sample height due to the tilt of the sample surface normal out of the plane defined by incoming and diffracted beams. In addition, $\omega$-$2\theta$ scans with a divergent beam Bragg-Brentano geometry were acquired over a large $2\theta$ range from 5-85° in order to detect small inclusions of possible secondary phases or misoriented grains.

**III. Results and Discussion**

The Mg-content $x$ in the $Ti_{1-x}Mg_xN$ layers measured by XPS is zero for pure TiN and increases to $x$ = 0.12, 0.19, 0.26, 0.39, 0.45 and 0.49 for deposition with an increasing power to the Mg target $P_{Mg}$ = 20, 40, 65, 100, 150 and 350 W while keeping the Ti flux constant. We note that $x$ increases less than linearly with $P_{Mg}$. This is most pronounced at high $x$ where an increase in $P_{Mg}$ by more than a factor of two from 150 to 350 W leads to only a 9% increase in $x$ from 0.45 to 0.49. This is attributed to a reduced Mg incorporation and Mg loss due to evaporation from the sample surface during deposition for layers with a large Mg content [13]. RBS compositional analyses for the same sample set yields $x$ = 0.07, 0.14, 0.16, 0.40, 0.43 and 0.45. That is, they agree to within 8% with the XPS measurements for samples with $P_{Mg} \geq 100$ W but underestimate the Mg-content for $P_{Mg} \leq 65$ W. This is caused by the challenge in uniquely fitting RBS data due to the Mg signal from the MgO substrate which results in a particularly large uncertainty for layers with small Mg content. Thus, in the remainder of the paper, we use $x$ values determined by XPS which we believe is more accurate in determining the Mg-to-Ti ratio. In contrast, RBS is more accurate in measuring the N-to-metal ratio, because XPS is strongly affected by N-containing surface contaminations and sputter cleaning is ineffective as it causes considerable compositional deviations due to preferential sputtering [56, 57]. The N-to-metal ratio in the Ti$_{1-x}$Mg$_x$N layers determined by RBS ranges from 0.96 ± 0.04 for $x \leq 0.39$ to 0.90 ± 0.04 for $x \geq 0.45$, indicating that most layers are stoichiometric within experimental uncertainty, but that the layers with the highest Mg-content may be N deficient. This is consistent with expectations since the most stable pure magnesium nitride phase (Mg$_3$N$_2$) has a N-to-metal ratio of 0.67. It is also consistent with previous first-principles simulations [12] and experimental data [13] suggesting an under-stoichiometric N concentration for Mg-rich layers. We note however, that the experimental uncertainty in determining the N-to-metal ratio is relatively large, due to the low RBS scattering cross section for light elements like N. We discuss in the following the results from two sets of samples, as also mentioned in Section II. More specifically, the first set of relatively thin (≤ 58 nm) layers have thicknesses determined by XRR of $d$ = 54.4, 49.0, 43.3, 52.0, 57.5, 44.5 and 35.0 nm for $P_{Mg}$ = 0, 20, 40, 65, 100, 150 and 350 W, respectively, and compositions as stated above. The second set of samples have larger thicknesses ($d \geq 110$ nm) which are measured by scanning electron microscopy as previously reported in Ref. [13] and are $d$ = 110, 150, and 275 nm for $x$ = 0.00, 0.37, and 0.49, respectively.

Figure 1(a) shows sections of typical x-ray diffraction $\omega$-$2\theta$ patterns over the $2\theta$ = 41.4°-43.5° range from 35.0 – 57.5 nm thick Ti$_{1-x}$Mg$_x$N layers with $x$ = 0-0.49, as labeled. The intensity is plotted on a logarithmic scale and the scans are offset vertically for clarity purposes. All patterns exhibit a Ti$_{1-x}$Mg$_x$N 002 peak at $2\theta$ = 42.1° - 42.5° and a MgO 002 substrate peak at $2\theta$ = 42.91°, corresponding to a lattice constant $a_o$ = 4.212 Å which is in agreement with the published value from bulk MgO crystals [58]. All samples exhibit only these two peaks over the entire measured $2\theta$ = 10°-85° range, indicating strong preferred 001 orientation of the Ti$_{1-x}$Mg$_x$N layers. The pattern from the TiN layer ($x$ = 0.00) has a peak at 42.50°, corresponding to an out-of-plane lattice constant $a_\perp$ = 4.251 Å. This value is within the range of previously reported values of 4.233-4.277 Å for epitaxial TiN/MgO(001) layers [8, 19-21, 41, 59, 60], with the majority of values within a narrower range of 4.240-4.249 Å. We attribute the relatively large out-of-plane lattice constant here to residual biaxial compressive stress which is affected by the growth temperature and layer thickness, as discussed in detail below. The pattern also exhibits Laue oscillations (layer thickness fringes) which indicate a relatively high crystalline quality and provide a value for the TiN layer thickness of 55 nm, determined using $<D>_{Laue} = \lambda/(2\sin\theta_{i+1} - 2\sin\theta_i)$, where $\lambda$ is the x-ray wavelength and $\theta_{i+1}$ and $\theta_i$ are angles of two adjacent maxima or minima [61]. This thickness value is in good agreement with $d$ = 54.4 nm obtained from XRR measurements. The alloy film with $x$ = 0.12 yields a similar XRD pattern as for pure TiN. The layer peak is slightly shifted to the left to $2\theta = 42.47°$, corresponding to a lattice parameter in the growth direction of $a_\perp = 4.253$ Å, and the Laue oscillations indicate a thickness of 47 nm, in reasonable agreement with 49.0 nm from XRR. Increasing the Mg concentration further to $x = 0.19, 0.26, 0.39, 0.45$ and $0.49$ leads to a continuous peak shift to smaller $2\theta$-values of $42.41°$, $42.36°, 42.25°, 42.16°$, and $42.10°$, corresponding to an increasing $a_\perp = 4.259, 4.264, 4.274, 4.284$, and $4.289$ Å, respectively. This increase is expected based on the predicted lattice constant of rock-salt structure MgN (4.34-4.44 Å) which is 2-4% larger than that for TiN [62], as discussed more quantitatively below. The Laue oscillations are well developed for $x \leq 0.26$, have a reduced amplitude for $x = 0.39$, and cannot be resolved for $x = 0.45$ and $0.49$. In addition, the layers with a large Mg content also have a broader and less intense $Ti_{1-x}Mg_xN$ 002 peak. More specifically, the peak intensity from layers with $x = 0.39, 0.45$ and $0.49$ is 2.7, 17 and 41 times weaker than that for pure TiN, respectively, and the peak full-width at half-maximum (FWHM) increases from $\Gamma_{2\theta} = 0.15°$ for $x = 0.39$ to $\Gamma_{2\theta} = 0.19°$ and $0.32°$ for $x = 0.45$ and $0.49$, indicating a decreasing crystalline quality which is attributed to a decreasing phase stability as the Mg-to-Ti ratio approaches unity [13].

Figure 1(b) shows a typical XRD $\omega$ rocking curve for the 002 reflection from the $Ti_{0.55}Mg_{0.45}N$ layer, which is obtained using a constant $2\theta = 42.16°$. Its FWHM of $\Gamma_\omega = 0.08°$ is quite small in comparison to other epitaxial rock-salt structure nitrides with reported values ranging from $0.14 – 2.20°$ [21, 23, 28-30, 34, 38, 42], indicating strong crystalline alignment of the $Ti_{0.55}Mg_{0.45}N$ 001 planes. The rocking curve width remains relatively small for all layers with $x \leq 0.45$, with $\Gamma_\omega = 0.06°, 0.07°, 0.04°, 0.03°, 0.06°$, and $0.08°$ for $x = 0, 0.12, 0.19, 0.26, 0.39$, and $0.45$, respectively. The smallest $\Gamma_\omega$ value in this list is from the $Ti_{0.74}Mg_{0.26}N$ layer which correspondingly also exhibits the largest peak intensity in Fig. 1(a). In contrast, the $Ti_{0.51}Mg_{0.49}N$ layer has a much larger $\Gamma_\omega = 0.50°$, indicating lattice distorting strain fields in this layer which has a composition close to the phase separation limit [12, 13], which is attributed to rock-salt structure MgN not being a thermodynamically stable phase such that an increasing $x$ causes compositional variations and/or the nucleation of alternative phases. The general trend of a decreasing crystalline quality for Mg rich layers is also consistent with the increasing $\Gamma_{2\theta}$ shown in Fig. 1(a).

Figure 1(c) shows a typical XRD $\varphi$-scan of the asymmetric 111 reflections from a $Ti_{0.55}Mg_{0.45}N$ layer. It is acquired using a $54.74°$ offset in $\chi$ and fixed $2\theta = 36.30°$ and $\omega = 18.15°$ to detect 111 reflections of the $Ti_{0.55}Mg_{0.45}N$ layer by recording the reflected intensity as a function of the azimuthal angle $\varphi$. The plot shows four equally spaced peaks labeled as 111, $\bar{1}11$, $1\bar{1}1$, and $\bar{1}\bar{1}1$, indicating that the layer has a single in-plane orientation. Fig. 1(c) also includes a $\varphi$-scan from the same sample, however, with $\omega$ and $2\theta$ adjusted to detect the substrate MgO 111 reflections. This scan exhibits four peaks that are $10^2$ times more intense but occur at the same $\varphi$ values as for the $Ti_{0.55}Mg_{0.45}N$ 111 reflections. This shows, in combination with the results from Figs. 1(a) and (b), that the $Ti_{0.55}Mg_{0.45}N$ layer exhibits a cube-on-cube epitaxial relationship with the substrate: $(001)_{TiMgN} \| (001)_{MgO}$ and $[100]_{TiMgN} \| [100]_{MgO}$. Similar XRD $\varphi$-scans were obtained also for the other samples, confirming that all layers in this study are epitaxial single crystals.

Figure 2 shows three typical high-resolution XRD reciprocal space maps acquired about asymmetric 113 reflections from $Ti_{1-x}Mg_xN(001)/MgO(001)$ layers with $x = 0.00, 0.26$, and $0.39$ and thickness $d = 54.4, 52.0$, and $57.5$ nm, respectively. The plot shows color filled iso-intensity contour maps in a logarithmic scale, plotted within $k$-space where $k_\perp = 2\sin\theta\cos(\omega-\theta)/\lambda$ and $k_\| =$ $2\sin\theta\sin(\omega-\theta)/\lambda$ correspond to directions perpendicular and parallel to the substrate surface along perpendicular [001] and [110] directions, respectively, as indicated by the labeled arrows. The plot also includes a scale bar in units of reciprocal length and arrows indicating the experimental $\omega$-$2\theta$ and $\omega$ scan directions. The strong peaks labeled "MgO" are due to MgO 113 substrate reflections, while the lower peaks are from $Ti_{1-x}Mg_xN$ 113. Some of the MgO 113 reflections indicate double or triple peaks and/or shoulders which originate from different domains in the substrates and are common for commercially available magnesium oxide single crystals [63]. We also note that the broadening of the MgO 113 peak along the $\omega$-direction is associated with small-angle grain boundaries in the single crystal substrates while the elongation from the top-left toward the bottom-right are intrinsic resolution streaks coming from the Ge(220) crystals in the hybrid monochromator [64].

The out-of-plane $a_\perp$ and in-plane $a_\parallel$ lattice constants are determined from the peak positions with $k_\perp = 7.122$ nm$^{-1}$ and $k_\parallel = 3.357$ nm$^{-1}$ for the MgO substrate and $k_\perp = 7.053$ nm$^{-1}$ and $k_\parallel = 3.358$ nm$^{-1}$ for the pure TiN ($x = 0.00$) layer, yielding $a_\perp = a_\parallel = 4.212 \pm 0.001$ Å for MgO and $a_\perp = 3/k_\perp = 4.254 \pm 0.001$ Å and $a_\parallel = \sqrt{2}/k_\parallel = 4.211 \pm 0.001$ Å for the TiN layer. Those values are consistent with the out-of-plane lattice parameters of $4.212 \pm 0.001$ Å and $4.251 \pm 0.001$ Å for MgO and TiN respectively, determined from $\omega$-$2\theta$ scans shown in Fig. 1. The TiN in-plane lattice parameter matches, within the experimental uncertainty, the MgO lattice constant. This indicates that the TiN layer is fully strained, that is, its lattice is coherent with that of the substrate, which is also directly evident from the vertical alignment of the MgO and TiN 113 reflections in Fig. 2(a), indicating the same $k_\parallel$ value for MgO and TiN. The elliptical shape of the TiN 113 reflection in reciprocal space can be described by peak broadening along the two primary directions $k_\perp$ and $k_\parallel$, which are rotated by 25.24° from the experimental $\omega$-$2\theta$ and $\omega$ scanning directions. The broadening along $k_\perp$ is attributed to local strain variations along the growth direction as well as the finite layer thickness, while broadening along $k_\parallel$ is primarily due to the mosaic spread, that is, the misalignment of crystallites which are single crystal blocks within the epitaxial TiN layer. The full-width at half-maximum peak widths are $\Delta k_\perp = 0.018$ nm$^{-1}$ and $\Delta k_\parallel = 0.006$ nm$^{-1}$. These values are in excellent agreement with $\Delta k_\perp = \Gamma_{2\theta}\cos\theta/\lambda = 0.017$ nm$^{-1}$ and $\Delta k_\parallel = 2\Gamma_\omega\sin\theta/\lambda = 0.005$ nm$^{-1}$, determined using the peak width $\Gamma_{2\theta}$ and $\Gamma_\omega$ from the symmetric $\omega$-$2\theta$ and $\omega$ rocking curve scans of the TiN 002 reflection (Fig. 1).

The reciprocal space maps in the middle and right of Fig. 2 are from a $Ti_{0.74}Mg_{0.26}N$ and a $Ti_{0.61}Mg_{0.39}N$ layer, respectively. Their $Ti_{1-x}Mg_xN$ 113 peaks are straight below the MgO 113 substrate peaks. This vertical alignment indicates that these layers are coherent with the substrate, that is, they are fully strained with $a_\parallel = a_{MgO}$, like for the case of the pure TiN layer. However, their $Ti_{1-x}Mg_xN$ 113 peaks are shifted downwards, indicating an increasing out-of-plane lattice constant $a_\perp$ which is attributed to an increasing relaxed lattice constant with increasing $x$, as discussed below. The $Ti_{1-x}Mg_xN$ 113 peaks are vertically elongated, which is primarily attributed to the finite layer thickness of 52.0 nm and 57.5 nm for the $Ti_{0.74}Mg_{0.26}N$ and the $Ti_{0.61}Mg_{0.39}N$ layers, respectively. In addition, particularly the peak at the right bottom in Fig. 2 from the $Ti_{0.61}Mg_{0.39}N$ 113 reflection is also elongated along the $\omega$ direction. This is attributed to the mosaicity of the substrate, as evident by the MgO 113 substrate reflection from this sample, showing a prominent multi-peak feature. The reciprocal space maps for all seven samples with $x = 0.0$-$0.49$ and $d = 35.0$-$57.5$ nm are qualitatively similar to those shown in Fig. 2. They indicate fully strained $Ti_{1-x}Mg_xN$ layers and show a continuous downwards shift of the $Ti_{1-x}Mg_xN$ 113 peak with increasing $x$, indicating an increasing $a_\perp$, as summarized further below.

Reciprocal space maps from considerably thicker epitaxial Ti$_{1-x}$Mg$_x$N(001) layers with $d$ = 110, 150, and 275 nm and $x$ = 0.00, 0.37, and 0.49 are shown in Figure 3. The dashed lines indicate 113 directions which extended from the origin to the upper right in the plot and correspond to the experimental $\omega$-$2\theta$ directions. Therefore, peaks which are centered on this line emanate from a perfectly cubic crystal with identical perpendicular and parallel lattice constants ($a_\perp = a_\parallel$). All MgO peaks are on the 113 lines, consistent with strain-free substrates, as expected. In contrast, the center of the 113 peak from the TiN ($x$ = 0) layer in the left panel of Fig. 3 is slightly to the right of the 113 line, indicating that the layer is strained with $a_\perp > a_\parallel$. More specifically, the peak position provides values for $a_\parallel$ = 4.219 ± 0.008 Å and $a_\perp$ = 4.243 ± 0.006 Å, suggesting a mild compressive strain $\varepsilon_\parallel = (a_\parallel/a_o - 1)$ = -0.34%, where the relaxed lattice constant $a_o$ is determined using $a_o = (a_\perp - va_\perp + 2va_\parallel)/(1 + v)$ = 4.234 Å and $v$ is the Poisson's ratio which is assumed to be a composition-independent constant of 0.22 [46, 60]. This assumption has a negligible effect on the plotted data since a change in $v$ from 0.22 to, for example, 0.25 causes a correction in $a_o$ of only 0.02%. In addition, the TiN 113 peak is to the left of the MgO 113 peak, indicating that $a_\parallel > a_{MgO}$. This is distinctly different from the map shown in Fig. 2 from the thinner TiN(001) layer with $d$ = 54.4 nm for which the 113 peak is vertically aligned with the substrate peak, indicating a fully strained layer with $a_\parallel = a_{MgO}$. Thus, the 110-nm-thick TiN(001) layer is partially relaxed, where the degree of relaxation $R = (a_\parallel - a_{MgO})/(a_o - a_{MgO})$ = 31 %. We note that the uncertainties in $a_\parallel$ and $a_\perp$ are larger than for the $d$ = 54.4 nm layer (Fig. 2), which is due to the difficulty in accurately determining the center of the 113 reflection in the reciprocal space map. The largest peak broadening is along the $\omega$ rocking-curve direction, which is primarily attributed to the mosaic spread, that is, the misalignment of the single crystal blocks within the TiN layer. The peak width $\Delta\omega$ = 0.27° can also be used to estimate an upper limit for the total dislocation density $\rho_{disl} \approx \Delta\omega^2/(2\pi\ln2 \times b^2)$ = 6 × 10$^9$ cm$^{-2}$ assuming uncorrelated (randomly distributed) dislocations [65, 66] where $b = a_{TiN}/\sqrt{2}$ = 0.300 nm is the Burgers vector along close-packed <110> directions in TiN.

The maps from the Ti$_{1-x}$Mg$_x$N alloy layers with $x$ = 0.37 and 0.49 in the middle and right panels of Fig. 3 show that both the substrate and layer peaks are centered on the dashed 113 lines. That is, these layers are fully relaxed ($a_\perp = a_\parallel$) within the experimental uncertainty. Analyzing the peak positions yields $a_\parallel$ = 4.254 ± 0.005 Å and $a_\perp$ = 4.259 ± 0.005 Å for the Ti$_{0.63}$Mg$_{0.37}$N layer and $a_\parallel$ = 4.268 ± 0.005 Å and $a_\perp$ = 4.275 ± 0.005 Å for the Ti$_{0.51}$Mg$_{0.49}$N layer. The $a_\perp$ values here agree well with 4.260 and 4.279 Å determined from XRD symmetric $\omega$-$2\theta$ scans for layers with $x$ = 0.37 and 0.49, respectively [13]. These two peaks are broader than the 113 peak from the TiN layer, indicating a larger mosaic spread. The measured $\Delta\omega$ = 0.60° and 0.73° yields an estimated upper bound for the dislocation density of 3 × 10$^{10}$ and 4 × 10$^{10}$ cm$^{-2}$ for $x$ = 0.37 and 0.49, respectively. These values are approximately five times larger than for the TiN ($x$ = 0) layer, which we attribute to the larger lattice misfit with the substrate and a correspondingly larger misfit dislocation density. The increasing misfit is evident in Fig. 3 from the increasing distance between layer and substrate peaks with increasing $x$ = 0 to 0.37 to 0.49, indicating an increasing relaxed lattice constant for Ti$_{1-x}$Mg$_x$N alloys, as discussed in the following.

Figure 4 is a plot of the out-of-plane $a_\perp$, in-plane $a_\parallel$, and relaxed $a_o$ lattice constants vs $x$ = 0.0-0.49 in Ti$_{1-x}$Mg$_x$N, as determined from XRD RSMs and $\omega$-$2\theta$ scans from epitaxial layers with $d$ = 35-58 nm represented by circles and diamonds, and $d$ = 110-275 nm represented by triangles. The plot includes two datasets for $a_\perp$. The open diamond symbols indicate the values obtained from the 002 peak position in $\omega$-$2\theta$ scans as shown in Fig. 1(a), while the solid circles are determined from the $k_\perp$ values of the 113 reflection measured using RSMs such as those shown in Fig. 2. There is good overall agreement between these two distinct measurements from the same sample set with deviations of only 0.01-0.15%. The $a_\perp$ values increase with increasing $x$, indicating an increasing lattice constant with increasing Mg content, as mentioned above and discussed below. In contrast, the plotted $a_\parallel$ is independent of $x$ within the experimental uncertainty, ranging from 4.211-4.214 Å. That is, the measured $a_\parallel$ is identical (± 0.03% variation) to the lattice constant of the substrate $a_{MgO}$ = 4.211 Å, consistent with these layers being fully strained. The in-plane biaxial compressive strain $\varepsilon_\parallel = (a_\parallel/a_o)-1$ increases from -0.64% for TiN, to $\varepsilon_\parallel$ =-0.65%, -0.69%, -0.78%, -0.85%, -1.1%, and -1.3% for layers with $x$ = 0.12, 0.19, 0.26, 0.39, 0.45, and 0.49, respectively.

The plotted blue circles in Fig. 4 represent the $a_o$ values of the fully strained 35-58 nm thick layers, while the triangles are from the $d$ = 110-275 nm layers. The $a_o$ continuously increases from 4.238 Å for TiN to $a_o$ = 4.241, 4.242, 4.247, 4.250, 4.258 and 4.265 Å for the thin layers with $x$ = 0.12, 0.19, 0.26, 0.39, 0.45 and 0.49, respectively, and from 4.234 to 4.257 to 4.273 Å for the thicker layers with $x$ = 0, 0.37 and 0.49. The two data sets are in good agreement, with small deviations of 0.11-0.24 % which approximately match the experimental uncertainty associated with both XRD measurements and compositional analyses. They indicate a positive curvature of $a_o$ vs $x$. This trend is quantified by fitting $a_o$ with a bowing expression $a_o = (1-x)a_{TiN} + xa_{MgN} - bx(1-x)$, as shown as dashed line in Fig. 4, indicating a negative deviation from Vegard's law as previously reported for Ti$_{1-x}$Mg$_x$N [13]. The fitting yields a bowing parameter $b$ = 0.113 Å, and lattice constants for the binaries of $a_{TiN}$ = 4.239 Å and $a_{MgN}$ = 4.345 Å. The former value is in good agreement with 4.2386 and 4.2397 Å, the reported lattice parameter from epitaxial TiN(001)/MgO(001) deposited by reactive magnetron sputtering at low temperature and 700 °C, respectively [48, 60], while our extrapolated value of 4.345 Å for the rocksalt-structure MgN is in the range 4.34-4.46 Å from theoretical predictions [12, 62]. The plotted $a_o$ from the Ti$_{1-x}$Mg$_x$N alloys with $0 \leq x \leq 0.49$ are 0.4 to 1.2 % smaller than predictions from first principles calculations [12], which is primarily attributed to overestimated lattice constants in the simulations caused by the generalized gradient approximation. Our values are also 0.3 to 0.6% smaller than the out-of-plane lattice constants reported by Fenker *et al* [1], and 0.5 to 0.9% smaller than the values by Balzer *et al* [2], indicating that the polycrystalline Ti$_{1-x}$Mg$_x$N layers in these studies may have been in a state of compressive strain.

Figure 5 shows plots of the out-of-plane $\xi_\perp$ and in-plane $\xi_\parallel$ x-ray coherence lengths as a function of $x$ in Ti$_{1-x}$Mg$_x$N(001) layers. The plotted coherence lengths are from both sample sets, distinguished by red and blue symbols for thin and thick layers, respectively, and determined from symmetric 002 and asymmetric 113 reflections indicated by circles and diamonds. More specifically, the full-width at half-maximum peak intensities $\Gamma_\omega$ and $\Gamma_{2\theta}$ from Ti$_{1-x}$Mg$_x$N 002 reflections along the $\omega$ rocking curve and $\omega$-$2\theta$ scan directions yield coherence lengths $\xi_\parallel = \lambda/(2\Gamma_\omega \sin\theta)$ and $\xi_\perp = \lambda/(\Gamma_{2\theta} \cos\theta)$, while the peak widths $\Delta k_\parallel$ and $\Delta k_\perp$ of Ti$_{1-x}$Mg$_x$N 113 reflections in the reciprocal space maps yield alternative coherence length values determined using $\xi_\parallel = 1/\Delta k_\parallel$ and $\xi_\perp = 1/\Delta k_\perp$. The two independent methods are in excellent agreement, as indicated by the overlapping circles and diamonds, with deviations of 0-15% for all layers with $x \leq 0.39$ but larger deviations for $x$ = 0.45 and 0.49 due to the experimental uncertainty (see error bars) associated with weaker x-ray diffraction peaks for the layers with a large Mg content.

The plotted coherence length along the growth direction $\xi_\perp$ from the fully stained layers ($d \leq 58$ nm) has values ranging from 30-63 nm and nearly perfectly replicates the layer thickness which is also plotted in Fig. 5(a) as open triangles. This shows that $\xi_\perp$ is limited by the layer thickness and that effects from possible crystalline defects and strain variations along the growth direction are below the detection limit [50]. A quantitative comparison of $\xi_\perp$ from both methods and $d$ indicates that $\xi_\perp$ is slightly higher (4% to 22%) than $d$, except for the sample with the highest Mg-content $x = 0.49$, for which $\xi_\perp$ is 18% below $d$. This suggests a decreasing crystalline quality as $x$ approaches the critical composition for phase separation, as stated above and consistent with the decreasing XRD peak intensity in Fig. 1(a). This becomes more evident from the $\xi_\parallel$ vs $x$ plot for the strained layers: $\xi_\parallel$ shows no clear trend as a function of composition, with values ranging from 100-400 nm for $x \leq 0.45$, indicating no reduction in the crystalline quality as Mg is added to TiN. However, $\xi_\parallel$ drops by an order of magnitude to 30 nm for $x = 0.49$, suggesting a fast reduction in the distance between threading and misfit dislocations and confirming the deterioration of the crystalline quality as $x$ approaches 0.5. The x-ray coherence lengths of thick layers are considerably smaller than of the thin layers: $\xi_\perp$ [blue symbols in Fig. 5(a)] is 1.4, 3.2 and 7.9 times smaller than $d$ (blue open triangles) for $x = 0.0$, 0.37 and 0.49, respectively, while $\xi_\parallel$ = 44, 23, and 20 nm [blue symbols in Fig. 5(b)] are by up to an order of magnitude smaller than for the corresponding thin layers. This is directly attributed to the relaxation of the thick layers, which introduces misfit and associated threading dislocations which reduce the measured x-ray coherence lengths.

As a last point, we discuss the distinct differences between the thin ($d \leq 58$ nm) and thick ($d \geq 110$ nm) layers, that is, the transition from fully strained to relaxed layers with increasing $d$. Relaxation is facilitated by threading dislocation glide, which causes an increase in the average misfit dislocation line length as well as possible dislocation multiplications. It is well known that thin epitaxial layers are expected to grow fully coherent with the substrate if their strain energy is smaller than the energy of the misfit dislocations that would be required to relax the strain [67, 68]. The total elastic energy for a given strain is proportional to the film thickness $d$, resulting in an increasing driving force for strain relaxation with increasing $d$. Correspondingly, there is a critical minimum thickness $h_c$ for relaxation as defined by the Matthew-Blakeslee formula [69]

$$h_c = \frac{|\vec{b}|}{8\pi(1+\nu)\delta_f}\left[\ln\left(\frac{h_c}{|\vec{b}|}\right)+1\right], \tag{1}$$

where the magnitude of the Burgers vector along the close-packed <110> directions in Ti$_{1-x}$Mg$_x$N is $|\vec{b}| = a_o/\sqrt{2} = 3.00$ Å, and the Poission's ratio of Ti$_{1-x}$Mg$_x$N can be approximated as $\nu = 0.22$, based on the reported value for TiN [46, 60]. For pure TiN ($x = 0$), the lattice mismatch with the MgO substrate $\delta_f = a_{TiN}/a_{MgO} - 1 = 0.68\%$, calculated from the measured and reported room-temperature $a_{TiN} = 4.240$ Å [70] and $a_{MgO} = 4.2112$ Å [71]. This mismatch decreases with increasing temperature, reaching just 0.23% at the growth temperature $T_s = 600 \pm 50$ °C, as determined using the thermal expansion coefficients of $9.35\times10^{-6}$ K$^{-1}$ for TiN [70] and $13\times10^{-6}$ K$^{-1}$ for MgO [71]. Applying Eq. (1) yields then $h_c = 23$ nm for TiN/MgO at 600 °C. However, $h_c$ decreases to 5.6 nm when the sample is cooled to room temperature, consistent with a previously reported value of 6.5 nm [72]. Our TiN(001) layer with $d = 54.4$ nm is fully strained and the thicker layer with $d = 110$ nm is partially relaxed, as shown in Figs. 2 and 3. This suggests that the experimental $h_c$ for TiN is somewhere between 54 and 110 nm, which is larger than the $h_c$ estimates from Eq. (1). We attribute this disagreement to kinetic barriers for dislocation formation. More specifically, TiN grows fully coherent on MgO(001) at 600 °C, even if its thickness is approximately twice the estimated critical thickness $h_c = 23$ nm. During subsequent cooling, $h_c$ decreases to 5.6 nm which is 10 times smaller than $d$. Nevertheless, the

layer remains fully strained. We attribute this to the limited growth temperature (26% of the TiN melting point of 3222 K [20]) and the even lower temperature during cooling, which are insufficient to nucleate and grow misfit dislocations. This argument is consistent with reports showing that 300-nm-thick epitaxial TiN/MgO(001) layers are fully strained if deposited at low temperature (≤ 420 °C) using high-flux low-energy ion irradiation during growth [48], while TiN layers grown at elevated temperatures with equivalent thickness are either partially relaxed ($T_s$ = 700 °C) [60] or fully relaxed ($T_s$ = 700-850 °C) [48]. Increasing $x$ in Ti$_{1-x}$Mg$_x$N alloys from 0 to 0.49 causes an increase in the room-temperature lattice misfit with MgO from $\delta_f$ = 0.68% to 1.1%. Correspondingly, the estimate for $h_c$ decreases from 5.6 to 2.9 nm. These values are well below $d$ = 35-58 nm of all fully strained layers in our study. Thus, similar to the argument above, we attribute the suppression of relaxation to kinetic barriers for misfit dislocation formation and/or growth. In addition to the kinetic barriers present in TiN, the random distribution of cations in the Ti$_{1-x}$Mg$_x$N alloy layers and the associated strain fields may contribute to the suppression of easy glide of threading dislocations which, in turn, inhibits the growth of misfit dislocations and layer relaxation, despite the increasing strain energy with increasing $x$. Thus, in summary, the experimentally determined critical thickness for relaxation of epitaxial Ti$_{1-x}$Mg$_x$N(001)/MgO(001) is between 35 and 110 nm, which is 10-20 times larger than 2.9-5.6 nm estimated using the Matthew-Blakeslee formula. This is attributed to kinetic barriers where the limited growth/cooling temperatures and possible strain fields in the alloy delay/inhibit stress relaxation through dislocation nucleation and glide.

## IV. Conclusions

Ti$_{1-x}$Mg$_x$N layers deposited on MgO(001) by reactive magnetron co-sputtering at 600 °C are epitaxial single crystals with a cube-on-cube relationship with their substrates: (001)$_{TiMgN}$∥(001)$_{MgO}$ and [100]$_{TiMgN}$∥[100]$_{MgO}$. Layers with thickness 35-58 nm are fully strained, while layers with $d$ ≥ 110 nm are partially ($x$ = 0) or fully ($x$ = 0.37 and 0.39) relaxed. This indicates a critical thickness for relaxation between 35 and 110 nm, which is an order of magnitude larger than equilibrium predictions of 2.9-5.6 nm, suggesting kinetic barriers for misfit dislocation nucleation and/or growth that limits relaxation during layer growth at 600 °C and during subsequent cooling to room temperature. The relaxed lattice parameter determined from XRD RSMs is in good agreement between the thin and thick layers and increases according to $a_o$ = (1-$x$)$a_{TiN}$ + $xa_{MgN}$ − $bx$(1-$x$), where $a_{TiN}$ = 4.239 Å, $a_{MgN}$ = 4.345 Å, and the bowing parameter $b$ = 0.113 Å. The in-plane x-ray coherence length is large (100 nm ≤ $\xi_\parallel$ ≤ 400 nm) for 0.00 ≤ $x$ ≤ 0.45 but drops to $\xi_\parallel$ = 30 nm for $x$ = 0.49. Similarly, $\xi_\perp$ is 4-22% larger than $d$ for 0.00 ≤ $x$ ≤ 0.45 but 18% smaller than $d$ for $x$ = 0.49, indicating local strain variations and a decreasing crystalline quality as $x$ approaches the critical composition for phase separation. Similarly, layers with $d$ ≥ 110 nm also exhibit an approximately 10-fold reduced $\xi_\parallel$ which is attributed to strain relaxation that results in misfit and associated threading dislocations.

**Acknowledgements**

The authors acknowledge financial support by the National Science Foundation under Grant Nos. 1712752 and 1629230.

**Figures**

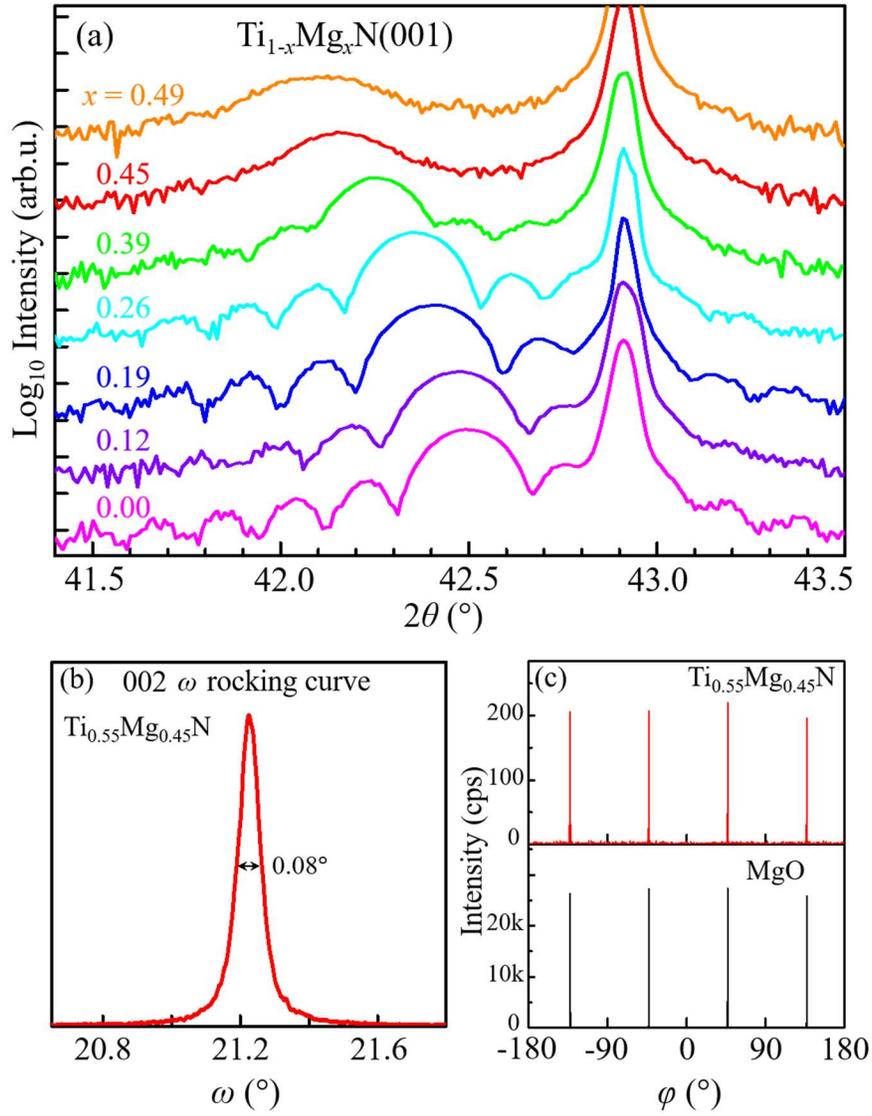

Figure 1: X-ray diffraction (a) $\omega$-$2\theta$ scans, (b) $\omega$ rocking curve of the 002 reflection, and (c) $\varphi$-scan of 111 reflections from $Ti_{1-x}Mg_xN(001)$ layers grown on MgO(001) with thickness $d$ = 35-58 nm.

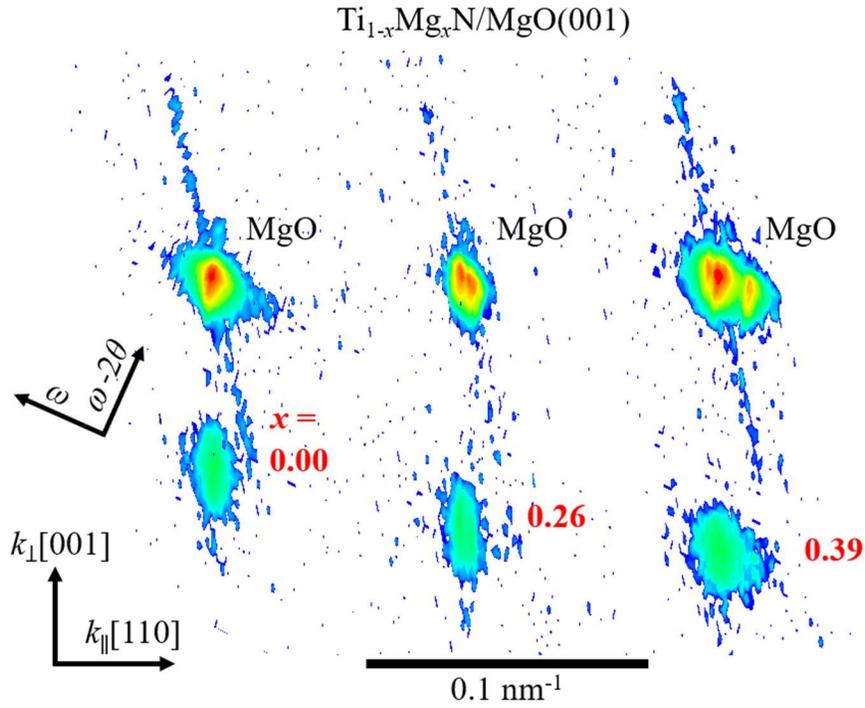

Figure 2: Representative RSMs about asymmetric 113 reflections from three 52 to 58 nm thick Ti$_{1-x}$Mg$_x$N(001) layers with $x$ = 0, 0.26 and 0.39.

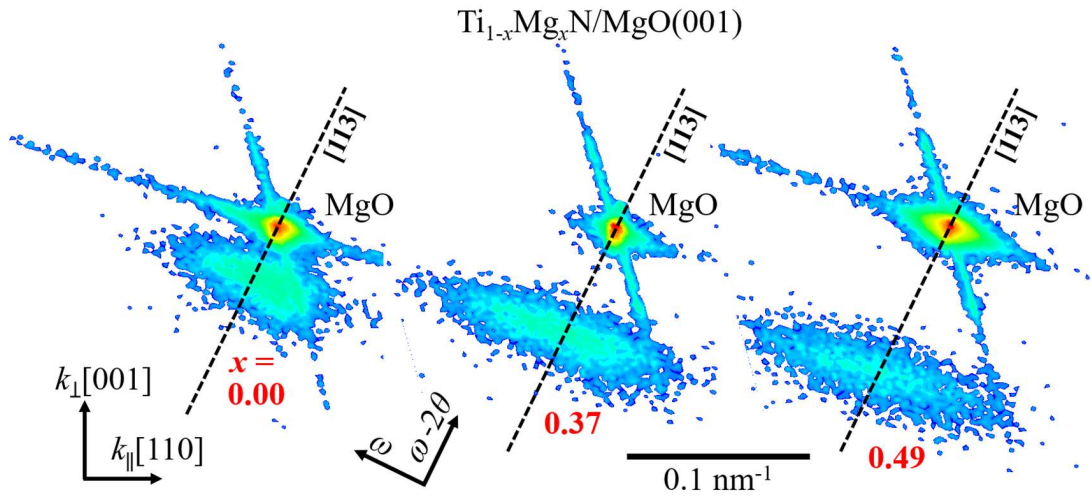

Figure 3: Representative RSMs about asymmetric 113 reflections from three ≥100 nm thick Ti$_{1-x}$Mg$_x$N(001) layers with $x$ = 0, 0.37 and 0.49.

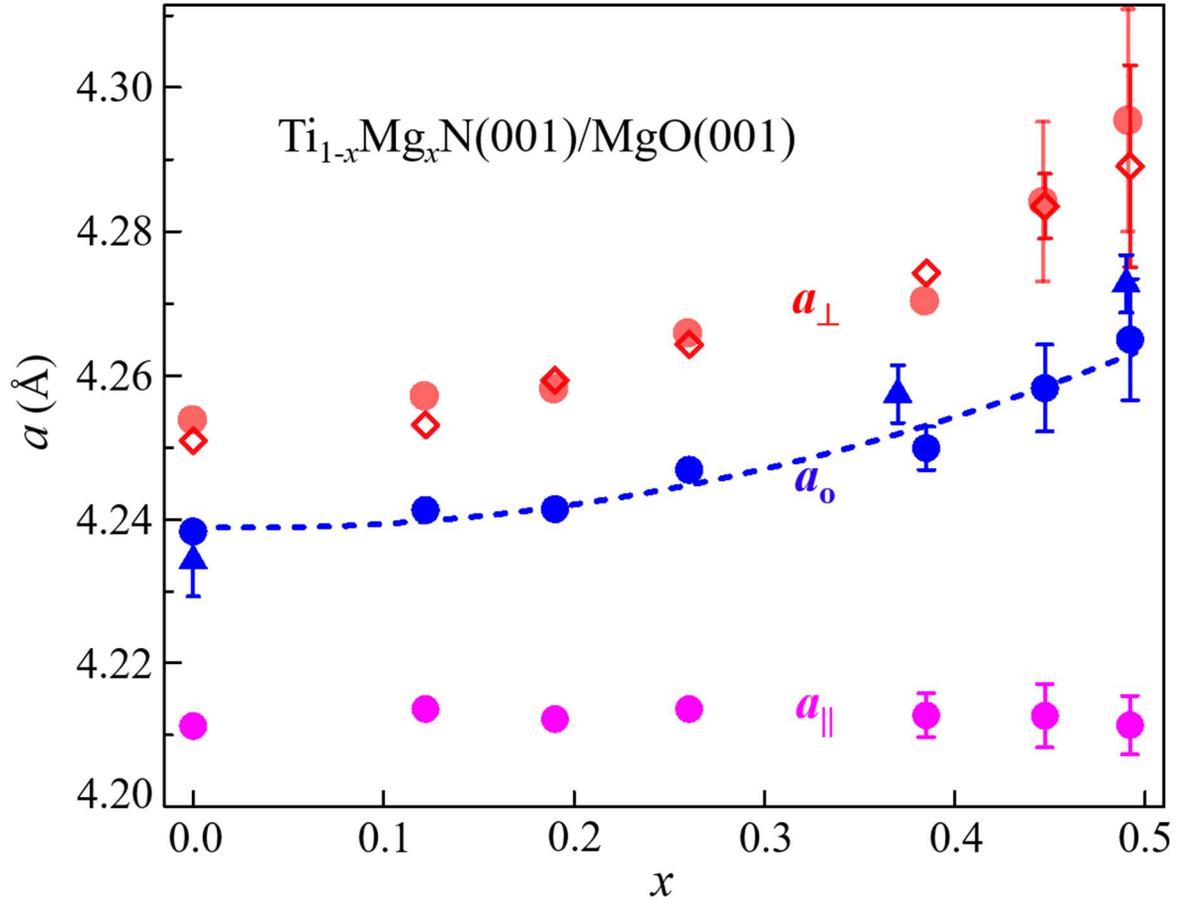

Figure 4: Out-of-plane $a_\perp$, in-plane $a_\parallel$, and relaxed $a_o$ lattice constants obtaining from RSMs (red, magenta and blue circles) and $a_\perp$ obtained from $\omega$-$2\theta$ scans (red open diamonds) vs Mg content $x$ in fully-strained Ti$_{1-x}$Mg$_x$N(001) layers grown on MgO(001) with $d \leq 58$ nm. The blue triangles show $a_o$ from relaxed Ti$_{1-x}$Mg$_x$N(001) layers with $d \geq 110$ nm.

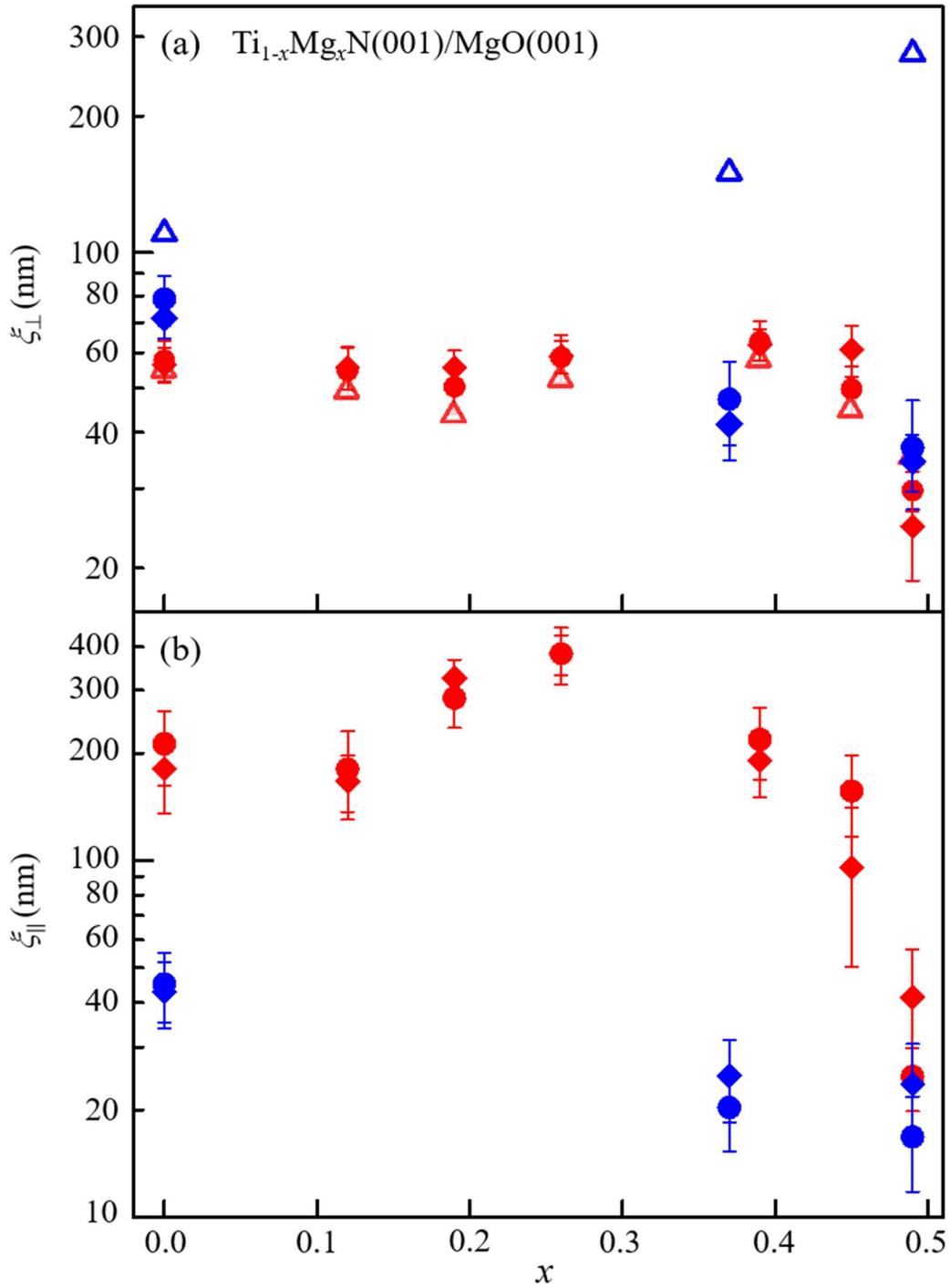

Figure 5: The (a) out-of-plane $\xi_\perp$ and (b) in-plane $\xi_\parallel$ x-ray coherence lengths from fully stained (red) and relaxed (blue) Ti$_{1-x}$Mg$_x$N(001) layers grown on MgO(001), determined from $\omega$-$2\theta$ and $\omega$ scans of 002 reflections (circles) and from RSMs of 113 reflections (diamonds). The layer thickness is indicated by triangles in (a).